\documentclass{kapproc} 

\kluwerbib
\normallatexbib

\usepackage{amsfonts}
\usepackage{epic}
                    
\def\cD{{\cal D}}

\def\cP{{\cal P}}                     
\def\cS{{\cal S}}          \def\cT{{\cal T}}          \def\cU{{\cal U}}

\def\ZZ{{\mathbb Z}}

\newcommand{\si}{\sigma}
\newcommand{\pr}{\prime}

\newcommand{\ot}{\otimes}
\newcommand{\nn}{\nonumber}

\begin{document}


\articletitle[Integrable chain models]
{Integrable chain models with\\
staggered R-matrices}

\author{Ara G. Sedrakyan}

\affil{Yerevan Physics Institute\\
Br.Alikhanian str.2, Yerevan 36, Armenia\\
and\\ 
Laboratoire d'Annecy-le-Vieux de Physique Th{\'e}orique LAPTH
\\
CNRS, UMR 5108, associ{\'e}e {\`a} l'Universit{\'e} de Savoie
\\
BP 110, F-74941 Annecy-le-Vieux Cedex, France}
\email{sedrak@lx2.yerphi.am;  sedrak@lappa.in2p3.fr}


\anxx{Sedrakyan\, Ara G.}

\begin{abstract}
The technique of construction on Manhattan lattice ($ML$)
the fermionic action for Integrable models is presented. The
Sign-Factor of 3D Ising model ($SF$ of $3DIM$) and Chalker-Coddington-s
phenomenological
model ($CCM$) for the edge excitations in Hall effect 
are formulated in this way. The second one demonstrates
the necessity to consider the inhomogeneous models with staggered
$R$-matrices. The disorder over the $U(1)$ phases is taken into
account and staggered Hubbard type of model is obtained. The technique
is developed to construct the integrable models with staggered
disposition of $R$-matrices.
\end{abstract}

\begin{flushright}
LAPTH-conf-879
\end{flushright}

\section{Introduction}

The goal of the talk is twofold. First is the demonstration 
how the integrable models, which can
be solved via the Bethe Ansatz ($BA$) technique and by definition
are in Hamiltonian formalism, can be formulated in the action
(Lagrangian) formalism on so called Manhattan lattice
($ML$) exactly.  More precisely  it will be shown how the partition function
of the model, which is a trace of $N$-degree of the Transfer matrix,
can be represented exactly as some functional integral over classical
Grassmann fields $\psi_{\vec{n}}$ with two dimensional action 
${\cal S}(\bar\psi_{\vec{n}};\psi_{\vec{n}})$
defined on the $ML$
\begin{equation}
\label{Z}
\ZZ=\cT r T^N = \int\cD\bar\psi_{\vec{n}} \cD\psi_{\vec{n}}
e^{-{\cal S}(\bar\psi_{\vec{n}};\psi_{\vec{n}})}.
\end{equation}

We start by demonstrating
that two interesting problems of modern physics, namely the
so called Sign-factor of three dimensional Ising model ($SF$ of $3DIM$)
\cite{P, FSS, KS} and the edge excitations in Hall effect (more
precisely the Chalker-Coddington ($CCM$) phenomenological model before
taking into account the disorder over random phases \cite{CC})
can be described by the same type of 2D model on $ML$ \cite{S}, 
but have a different amount of 
degrees of freedom (correspondingly different gauge groups of 
symmetries) and are in different points of the space of hopping
parameters. 

The formulation of $CCM$
as a field theory of scalar fermions on $ML$ in the $U(1)$ gauge field
background exhibits chess like structure and demonstrates
the necessity to consider and investigate
an inhomogeneous integrable models with staggered disposition
of the R-matrices along a chain and time directions. 

It is turning out, that this formalism on $ML$ is very appropriate
for taking into account the disorder over $U(1)$ phases
in the $CCM$  in order to analyze its Lyapunov index (which
defines the correlation length index for the edge excitations).
In a result the Hubbard type model with staggered disposition
of R-matrices is appearing.

In the action formalism also becomes evident how the models can be formulated
on the random $ML$, which will allow to develop the string model
corresponding to them.

In a second part of the talk the integrable models will be analyzed, 
where the Monodromy matrix is defined as a two row product of
staggered R-matrices. 
The corresponding
Yang-Baxter equations ($YBE$), which ensures the commutativity
of Transfer matrices of different values of the spectral parameter
will be presented \cite{ASSS1}. It appeared, that the modified $YBE$'s have a
solution for $U_q(sl(n))$ groups giving rise of the models with
staggered signs of the anisotropy parameter $\Delta$. Since in this 
construction the $R(u)$-matrices in the product has also staggered shift of the
spectral parameter $u$ by new model parameter $\theta$, as the
calculations of the Hamiltonian shows, they can be regarded
as a models on the zig-zag ladder chains. In the XXZ \cite{APSS},
anisotropic t-J \cite{TS} and Hubbard cases \cite{ASSS2} 
the Hamiltonian is found explicitly. 
The quantum group structure, which is behind of this construction in
the $sl(n)$ case was analyzed in the article \cite{ASSS1}.
\section[The $SF$ of $3DIM$ and Field Theory Formulation]
{The $SF$ of $3DIM$ and Field Theory Formulation\\ of $CCM$
on 2D $ML$}

In the article \cite{KS} the model for $SF$ of $3DIM$ \cite{P,FSS,KS}
was formulated on the random $ML$, which is induced by the random
closed surface in 3D regular lattice. But for simplicity we will
consider now the flat $ML$ and outlined the essential characteristics
of the model.

 The Manhattan lattice ($ML$) is the lattice, where there are
continuous arrows on the links with the opposite directions
on the neighbor parallel lines (Fig.1). The arrows form a set of 
vectors
$\vec{\mu}_{ij}\in {\cal S}$. $ML$ originally was defined by Kasteleyn
\cite{K} in connection with the problem of single Hamiltonian
 walk.

\begin{figure}[ht]
\begin{center}
\vskip.2in
\begingroup\makeatletter\ifx\SetFigFont\undefined%
\gdef\SetFigFont#1#2#3#4#5{%
  \reset@font\fontsize{#1}{#2pt}%
  \fontfamily{#3}\fontseries{#4}\fontshape{#5}%
  \selectfont}%
\fi\endgroup%
\small{
\setlength{\unitlength}{6mm}
\begin{picture}(15,13)
\multiput(0,2)(3,0){5}{\vector(1,0){3}}
\multiput(3,5)(3,0){5}{\vector(-1,0){3}}
\multiput(0,8)(3,0){5}{\vector(1,0){3}}
\multiput(3,11)(3,0){5}{\vector(-1,0){3}}
\multiput(1.5,0)(0,3){4}{\vector(0,1){3}}
\multiput(4.5,3)(0,3){4}{\vector(0,-1){3}}
\multiput(7.5,0)(0,3){4}{\vector(0,1){3}}
\multiput(10.5,3)(0,3){4}{\vector(0,-1){3}}
\multiput(13.5,0)(0,3){4}{\vector(0,1){3}} 
\put(3,3.5){\shortstack{\circle{0.2}}}
\put(3,6.5){\shortstack{\circle*{0.2}}}
\put(6,3.5){\shortstack{\circle*{0.2}}}
\put(6,6.5){\shortstack{\circle{0.2}}}
\put(9,3.5){\shortstack{\circle{0.2}}}
\put(9,6.5){\shortstack{\circle*{0.2}}}
\put(12,3.5){\shortstack{\circle*{0.2}}}
\put(12,6.5){\shortstack{\circle{0.2}}}
\put(3,9.5){\shortstack{\circle{0.2}}}
\put(6,9.5){\shortstack{\circle*{0.2}}}
\put(9,9.5){\shortstack{\circle{0.2}}}
\put(12,9.5){\shortstack{\circle*{0.2}}}
\put(2.5,9){\shortstack{$B_1$}}
\put(5.5,9){\shortstack{$A_2$}}
\put(8.5,9){\shortstack{$B_1$}}
\put(11.5,9){\shortstack{$A_2$}}
\put(3.8,8.3){\shortstack{2}}
\put(6.8,8.3){\shortstack{3}}
\put(9.8,8.3){\shortstack{2}}
\put(12.8,8.3){\shortstack{3}}
\put(3.8,5.3){\shortstack{1}}
\put(6.8,5.3){\shortstack{4}}
\put(9.8,5.3){\shortstack{1}}
\put(12.8,5.3){\shortstack{4}}
\put(3.8,2.3){\shortstack{2}}
\put(6.8,2.3){\shortstack{3}}
\put(9.8,2.3){\shortstack{2}}
\put(12.8,2.3){\shortstack{3}}
\put(2.5,3){\shortstack{$B_1$}}
\put(5.5,3){\shortstack{$A_2$}}
\put(8.5,3){\shortstack{$B_1$}}
\put(11.5,3){\shortstack{$A_2$}}
\put(2.5,6){\shortstack{$A_1$}}
\put(5.5,6){\shortstack{$B_2$}}
\put(8.5,6){\shortstack{$A_1$}}
\put(11.5,6){\shortstack{$B_2$}}
\put(3.8,-1){\shortstack{2j-1}}
\put(1,-1){\shortstack{2j-2}}
\put(7,-1){\shortstack{2j}}
\put(10,-1){\shortstack{2j+1}}
\put(-2,2){\shortstack{m}}
\put(-2,5){\shortstack{m+1}}
\end{picture}}
\end{center}
\caption{Manhattan lattice.}
\end{figure}
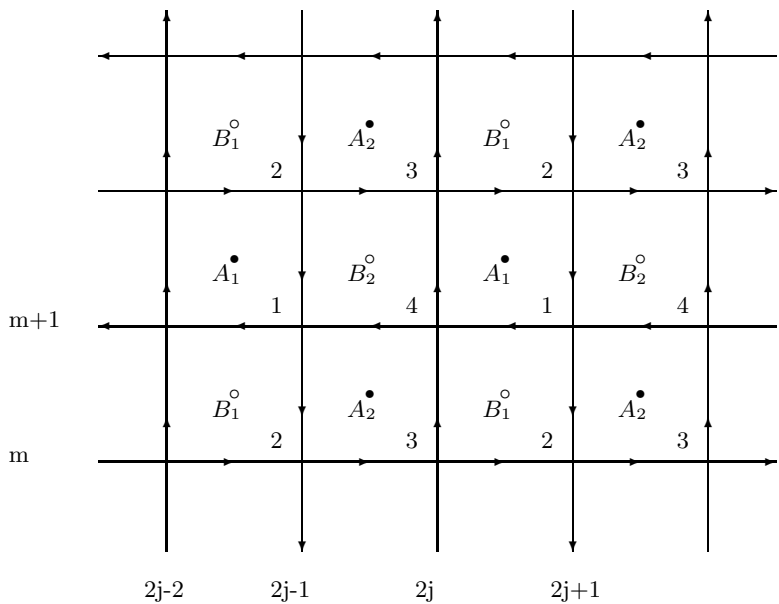
\vspace{15mm}

The plaquettes of $ML$ are divided into four groups, $A_a$ and $B_a$
 (a=1,2),
destinated in the chess like order. The A-plaquettes differ from
B-plaquettes by the fact, that arrows are circulating around them,
while there is no circulation for B-plaquettes. $A_1(A_2)$ has 
clockwise(counterclockwise) circulation, while $B_1$ differs from the
$B_2$ by rotation on $\pi/4$.

Consider the field of Grassmann variables 
$\Psi_{\vec{n}}= \left(\begin{array}{l}
\psi_{\vec{n},L}\nn\\
\psi_{\vec{n},L}\nn\end{array}\right)$ at the sites $\vec n$ of $ML$,
which is spinor irrep of $SO(3)$ (or fundamental irrep of $SU(2)$),
but forbid the double occupancy of all sites by fermions. This can be
acheved, for example, by putting the projectors $\Delta_{\vec{n}}=
\bar\psi_{\vec{n},L}\psi_{\vec{n},L}+\bar\psi_{\vec{n},R}\psi_{\vec{n},R}$
to the sites. Following the article \cite{KS} let us write the action
of this fields as for fermions, hoping only along arrows of $ML$
and being in the external $SU(2)$ gauge field, which is induced by
the immersion of the 2D surface into the 3D Euclidean space (see
details in \cite{KS}). Then this action defines the model for
$SF$ of $3DIM$.

In 1988 J. Chalker and P.D. Coddington \cite{CC} have defined a 
phenomenological model in the Transfer matrix formalism in order
to describe the edge excitations in Hall effect, responsible for
plateau-plateau transitions. Remarkably, the numerical simulations
give the desired experimental value for the correlation length
index, approximately (may be exactly) equal to 7/3.

We will see now, that if one will consider on $ML$ an action of
scalar Grassmann fields $\psi_{\vec n}$, which are hopping
in the $U(1)$ gauge field along arrows with appropriate hopping
parameters and by use of coherent states \cite{B, F} pass to Transfer
matrix (Hamiltonian) of discrete time evolution (as it is done
in \cite{S}), then in one particle sector the Transfer
matrix of $CCM$ before averaging over
random phases will be reproduced. The action of the model is
\begin{eqnarray}
\label{ACT}
-{\cal S}(\bar\psi_{\vec{n}};\psi_{\vec{n}})
= \sum_{\vec{n},\atop{\beta=1,2}}
t_{\vec n,\vec n+\vec{\mu}_\beta(\vec n)}\bar\psi_{\vec n} U_{\vec n}
\psi_{\vec n+\vec{\mu}_\beta(\vec n)}
+\sum_{\vec{n}}\bar\psi_{\vec n}\psi_{\vec n}.
\end{eqnarray}
In this expression $\vec{\mu_{\beta}(\vec n)}$, $\beta = 1,2$
are the fields of unit vectors on $ML$ defined at each site $\vec n$
and directed along two exiting arrows and 
$t_{\vec n,\vec n+\vec{\mu}_\beta(\vec n)}$ are the
hopping parameters between the points $\vec n$ and 
$\vec n+\vec{\mu}_\beta(\vec n)$. 
Because the structure of
$ML$ is translational invariant on two lattice spacing in
both(time and space) directions, which we would like to maintain,
in most general case one can consider only eight different 
hopping parameters . Below, in correspondence with notations on Fig.1,
 we will mark the hopping parameters from $j$ to
$i \;\;(i,j=1,2,3,4)$ as $t_{ij}$. 

The field of phase factors 
$U_{\vec n}= e^{i \alpha_{\vec n}}$
is independent of $\beta =1,2$(the phase factors on the exiting
from the site $\vec n$ two links are the same).
This distribution of phases on $ML$ is in exact correspondence with
$CCM$ and defines the $U(1)$- curvature equal to zero for the all
B-plaquettes, while random curvatures are located in the A-plaquetts.
It is also in clear correspondence with the random $ML$ picture
for the $SF$ of $3DIM$ \cite{KS}, where the curvatures,
induced by immersions of 2d surfaces in 3D regular lattice,
are located in the A-plaquettes.

Let us introduce now the fermionic coherent states according to
articles \cite{B, F} and pass to fermionic Transfer matrix
as it is done in \cite{S}.
\begin{eqnarray}
\label{CS1}
|\psi_{2j}\rangle = e^{\psi_{2j}c^+_{2j}}|0\rangle ,\qquad 
\langle\bar{\psi}_{2j}| = \langle0| e^{c_{2j}\bar{\psi}_{2j}}
\end{eqnarray}
for the even sites of the chain and
\begin{eqnarray}
\label{CS2}
|\bar{\psi}_{2j+1}\rangle = (c^+_{2j+1}-\bar{\psi}_{2j+1})
|0\rangle ,\qquad
\langle\psi_{2j+1}| = \langle 0|(c_{2j+1}+\psi_{2j+1})
\end{eqnarray}
for the odd sites. 

This states are designed as an eigenstates of
creation-annihilation operators of fermions $c_j^+,\;\; c_j $
with eigenvalues $\psi_j$ and $\bar\psi_j$
\begin{eqnarray}
\label{CPR}
c_{2j}\mid \psi_{2j}\rangle =- \psi_{2j}\mid \psi_{2j}\rangle &,&
\langle\bar{\psi}_{2j} \mid c^+_{2j} = 
-\langle\bar{\psi}_{2j} \mid \bar{\psi}_{2j},\\
c^+_{2j+1}\mid \bar{\psi}_{2j+1}\rangle =
\bar{\psi}_{2j+1}\mid \bar{\psi}_{2j+1}\rangle &,&
\langle\psi_{2j+1}\mid c_{2j+1} =- \langle\psi_{2j+1}\mid \psi_{2j+1}.\nn
\end{eqnarray}

It is easy to calculate the scalar product of this states
\begin{eqnarray}
\langle \bar{\psi}_{2j}\mid \psi_{2j}\rangle =
e^{\bar{\psi}_{2j} \psi_{2j}},\nn\\
\label{TpR}
\langle\psi_{2j+1} \mid \bar{\psi}_{2j+1}\rangle =
e^{\bar{\psi}_{2j+1} \psi_{2j+1}}
\end{eqnarray}

and find the completeness relations  
\begin{eqnarray}
\int d\bar{\psi}_{2j}d\psi_{2j}\mid \psi_{2j}
\rangle \langle\bar{\psi}_{2j}\mid e^{\psi_{2j}\bar{\psi}_{2j}}&=& 1,
\nn\\
\label{CRE}
\int d\bar{\psi}_{2j+1}d\psi_{2j+1}\mid \bar{\psi}_{2j+1}\rangle 
\langle\psi_{2j+1}\mid e^{\psi_{2j+1}\bar{\psi}_{2j+1}}&=& 1.
\end{eqnarray}

Let us attach the Fock spaces $V_j$ of scalar fermions $c_j^+, \;
c_j$ to each site of the chain and consider two type of $R$-matrices 
in the braid formalism in the operator form
\begin{eqnarray}
&&\check R_{2j,2j\pm1}=\nn\\
\label{XX}
&=&a_{\pm1}n_{2j}n_{2j\pm1} + a_{\pm2}(1-n_{2j})(1-n_{2j\pm1})+ 
n_{2j}(1-n_{2j\pm1})\\
&+&(a_{\pm1}a_{\pm2}+b_{\pm1}b_{\pm2})n_{2j\pm1}(1-n_{2j})
+b_{\pm1}c_{2j}^+c_{2j\pm1}+b_{\pm2}c_{2j\pm1}^+c_{2j}\nn\\
&=&:e^{\left[b_{\pm1}c_{2j\pm1}^+c_{2j} +b_{\pm2}c_{2j}^+c_{2j\pm1}
+(a_{\pm1}-1)c_{2j}^+c_{2j}+(1-a_{\pm2})c_{2j\pm1}^+c_{2j\pm1}\right]}:\nn
\end{eqnarray}
corresponding to two type of $B$-plaquettes on $ML$, with
\begin{eqnarray}
\label{ta}
a_{+1}=e^{i\alpha_3}t_{43},\;a_{+2}=e^{i\alpha_1}t_{21},\;
b_{+1}=-e^{i\alpha_3}t_{23},\;
b_{+2}=e^{i\alpha_1}t_{41},\; for \; B_1,\\
a_{-1}=e^{i\alpha_4}t_{34},\;a_{-2}=e^{i\alpha_2}t_{12},\;
b_{-1}=-e^{i\alpha_4}t_{14},\;
b_{-2}=e^{i\alpha_2}t_{32},\; for \; B_2.\nn
\end{eqnarray}
and where the symbol $: \;\;:$  means normal ordering
of fermionic operators in the even sites and anti-normal
(hole) ordering for the odd sites. This operators are acting
on the direct product of two neighbor Fock spaces
$V_{2j}\ot V_{2j\pm1}$ and are nothing, but the fermionized
versions of $R$-matrices of the ordinary $XX$ models
\begin{eqnarray}
\label{XXR}
\check{R}_{\pm} = \left(\begin{array}{llll}
a_{\pm1}&0&0&0\\
0&1&b_{\pm1}&0\\
0&b_{\pm2}&(a_{\pm1}a_{\pm2}+b_{\pm1})b_{\pm2}&0\\
0&0&0&a_{\pm2}
\end{array}\right),
\end{eqnarray}
which can be found by Jordan-Wigner transformation \cite{W}
or by the alternative technique, developed in \cite{AKMS}.
Considering now two Monodromy matrices $M_1$ and $M_2$ as a product
of $R$-matrices(corresponding to $B$-plaquettes) along the neighbor
rows
\begin{eqnarray}
\label{MRR}
M_1=\prod_j \check{R}_{2j,2j+1},\qquad M_2=\prod_j \check{R}_{2j,2j-1},
\end{eqnarray}
one can show that the Transfer matrix $T=\cT rM_1M_2$ defines
the partition function $\ZZ$ according to formulas (\ref{Z}) and 
(\ref{ACT}). Indeed, let us in the space of states $\prod_j V_j$ 
of the chain pass to the coherent basis and calculate the matrix elements
of the $R_{2j,2j\pm1}$-operators between the initial
$\mid \psi_{2j}\rangle ,\;\; \mid \bar{\psi}_{2j\pm1}\rangle$
and final $\langle\bar\psi_{2j}^{\pr}\mid,
\;\;\langle\psi_{2j\pm1}^{\pr}\mid$
states. By use of properties of coherent states it is easy to find
from the formula (\ref{XX}), that 
\begin{eqnarray}
R_{\psi_{2j},\bar{\psi}_{2j\pm1}}^{\bar{\psi}^{\pr}_{2j},\psi^{\pr}_{2j\pm1}}=
\langle\psi^{\pr}_{2j\pm1},\bar{\psi}^{\pr}_{2j}\mid\check{R}_{2j,2j\pm1}
\mid\psi_{2j},\bar{\psi}_{2j\pm1}\rangle=\nn\\
\label{Rpsi}
=e^{\left[a_{\pm1}\bar{\psi}^{\pr}_{2j}\psi_{2j}
+a_{\pm2}\bar{\psi}_{2j\pm1}\psi^{\pr}_{2j\pm1}
-b_{\pm1}\bar{\psi}_{2j\pm1}\psi_{2j} +
b_{\pm2}\bar{\psi}^{\pr}_{2j}\psi^{\pr}_{2j\pm1}\right]},
\end{eqnarray}
which, together with multiplication rules due to completeness relations
(\ref{CRE}), demonstrates the correctness of the formula (\ref{Z})
(see \cite{S} for details).

Let us consider now the one-particle sector of Fock space of the chain
\begin{eqnarray}
\label{OP}
\mid i\rangle= c^+_i\mid0\rangle,\qquad\qquad i=1,...2N
\end{eqnarray}
and calculate the matrix elements of the operators $M_0$ and
$M_1$ (\ref{MRR}) in this basis. Then after
parameterizing the hopping parameters as
\begin{eqnarray}
-t_{23}=t_{12}=t_{34}=t_{41}= 1/\cosh\theta,\nn\\
\label{hop}
-t_{14}=t_{43}=t_{32}=t_{21}= \tanh\theta
\end{eqnarray}
and unessential rescaling of $M_0M_1$ by factor $(t_{12}t_{21})^N$,
one easily can recover the $2N\; X\; 2N$ Transfer matrix by 
Chalker and Coddington before the averaging over phases, introduced
in the article \cite{CC}. In order to make the correspondence
totally obvious one should change in the Fig.1 the $B$-plaquettes
bye the act of scattering, as it is drown by dashed lines in
the Fig.2. 
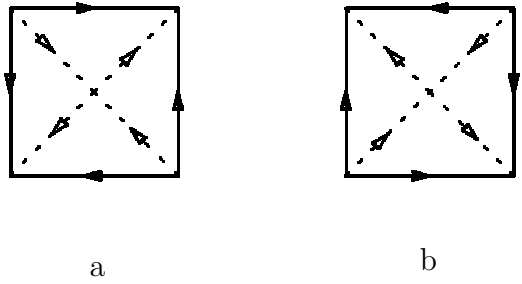
\begin{figure}[ht]
\begin{center}
\vskip.1in
\setlength{\unitlength}{0.000373333in}
\begingroup\makeatletter\ifx\SetFigFont\undefined%
\gdef\SetFigFont#1#2#3#4#5{%
  \reset@font\fontsize{#1}{#2pt}%
  \fontfamily{#3}\fontseries{#4}\fontshape{#5}%
  \selectfont}%
\fi\endgroup%
{\renewcommand{\dashlinestretch}{30}
\begin{picture}(7364,3922)(0,-10)
\thicklines
\drawline(82,3825)(1282,3825)
\drawline(1042.000,3765.000)(1282.000,3825.000)(1042.000,3885.000)(1042.000,3765.000)
\drawline(2482,1425)(2482,2625)
\drawline(2542.000,2385.000)(2482.000,2625.000)(2422.000,2385.000)(2542.000,2385.000)
\drawline(82,3825)(82,2625)
\drawline(22.000,2865.000)(82.000,2625.000)(142.000,2865.000)(22.000,2865.000)
\drawline(82,2625)(82,1425)
\drawline(2482,1425)(1132,1425)
\drawline(1372.000,1485.000)(1132.000,1425.000)(1372.000,1365.000)(1372.000,1485.000)
\drawline(1132,1425)(82,1425)
\dashline{90.000}(82,3825)(682,3225)
\drawline(469.868,3352.279)(682.000,3225.000)(554.721,3437.132)(469.868,3352.279)
\dashline{90.000}(682,3225)(1282,2625)
\dashline{90.000}(1282,2625)(1882,2025)
\dashline{90.000}(2482,1425)(1807,2100)
\drawline(2019.132,1972.721)(1807.000,2100.000)(1934.279,1887.868)(2019.132,1972.721)
\dashline{90.000}(1282,2625)(1882,3225)
\drawline(1754.721,3012.868)(1882.000,3225.000)(1669.868,3097.721)(1754.721,3012.868)
\dashline{90.000}(1282,2625)(682,2025)
\drawline(809.279,2237.132)(682.000,2025.000)(894.132,2152.279)(809.279,2237.132)
\dashline{90.000}(1882,3225)(2482,3825)
\dashline{90.000}(682,2025)(82,1425)
\drawline(6322.000,3885.000)(6082.000,3825.000)(6322.000,3765.000)(6322.000,3885.000)
\drawline(6082,3825)(7282,3825)
\drawline(1282,3825)(2482,3825)(2482,2550)
\drawline(7222.000,2865.000)(7282.000,2625.000)(7342.000,2865.000)(7222.000,2865.000)
\drawline(7282,2625)(7282,3825)
\put(5932,75){\makebox(0,0)[lb]{\smash{{{\SetFigFont{12}{13.0}{\rmdefault}{\mddefault}{\updefault}b}}}}}
\drawline(4882,1425)(4882,2625)
\drawline(4942.000,2385.000)(4882.000,2625.000)(4822.000,2385.000)(4942.000,2385.000)
\drawline(4882,1425)(6082,1425)
\drawline(5842.000,1365.000)(6082.000,1425.000)(5842.000,1485.000)(5842.000,1365.000)
\drawline(4882,2550)(4882,3825)
\drawline(4882,3825)(6157,3825)
\drawline(7282,2625)(7282,1425)
\drawline(6007,1425)(7282,1425)
\dashline{90.000}(4882,3825)(5482,3225)
\dashline{90.000}(7282,1425)(6682,2025)
\dashline{90.000}(4882,1425)(5482,2025)
\drawline(5354.721,1812.868)(5482.000,2025.000)(5269.868,1897.721)(5354.721,1812.868)
\dashline{90.000}(7282,3825)(6682,3225)
\drawline(6809.279,3437.132)(6682.000,3225.000)(6894.132,3352.279)(6809.279,3437.132)
\dashline{90.000}(6082,2625)(5482,3225)
\drawline(5694.132,3097.721)(5482.000,3225.000)(5609.279,3012.868)(5694.132,3097.721)
\dashline{90.000}(6082,2625)(6757,1950)
\drawline(6544.868,2077.279)(6757.000,1950.000)(6629.721,2162.132)(6544.868,2077.279)
\dashline{90.000}(5482,2025)(6682,3225)
\put(1207,0){\makebox(0,0)[lb]{\smash{{{\SetFigFont{12}{13.0}{\rmdefault}{\mddefault}{\updefault}a}}}}}
\end{picture}}
\end{center}
\caption{Scattering of particles in $CCM$ corresponding 
a) to $B_2$ - b) to $B_1$- plaquettes.}
\end{figure}

It appears that the action formalism by use of fermionic fields on
$ML$ is quite appropriate for taking now into account the disorder
over the $U(1)$ phases in the model and investigate, for example,
the Lyapunov index, which defines the correlation length index. For
Lyapunov index one should investigate the average over 
$\phi= -i\log U$ phases
of the square of the partition function $\langle \ZZ \ot \ZZ^+\rangle$
and we will consider the Gaussian distribution
$\cP(\{\phi_{\vec n}\})=
\prod_{\vec n}\frac{1}{\kappa\sqrt\pi}
exp\left(-\frac{\phi_{\vec n}^2}{\kappa^2}\right)$
for them. It is clear, that since phases are defined locally and
there is no correlations in averaging between different points,
we will have
\begin{eqnarray}
\label{ZZR}
\langle \ZZ \ot \ZZ^+\rangle=\cT r\left( \prod_j \langle 
\check{R}_{2j,2j+1}\ot\check{R}^+_{2j,2j+1}\rangle
\prod_i \langle 
\check{R}_{2i-1,2i}\ot\check{R}^+_{2i-1,2i}\rangle\right)^N
\end{eqnarray} 

The average $\langle\check{R}_{2j,2j\pm1}\ot\check{R}^+_{2j,2j\pm1}
\rangle$ is defining the $R$-operator of the new model and it is 
easy to calculate it in the $\psi$-basis of coherent states. Simple
Gaussian integration by use of expressions (\ref{Rpsi}) gives us
the $R$-matrix of the averaged model

\begin{eqnarray}
R_{\psi_{2j,\si},\bar{\psi}_{2j\pm1,\si}}^{\bar{\psi}^{\pr}_{2j,\si},
\psi^{\pr}_{2j\pm1,\si}}=
exp\left\{  \sum_{\si=\uparrow,\downarrow}
\left[\bar{a}_{\pm}(\bar{\psi}^{\pr}_{2j,\si}\psi_{2j,\si}
+\bar{\psi}_{2j\pm1,\si}\psi^{\pr}_{2j\pm1,\si})
\right.\right.\nn\\
\label{Rhub}
\left.+(-)^{\si}\bar{b}_{\pm}(-\bar{\psi}_{2j\pm1,\si}
\psi_{2j,\si} + \bar{\psi}^{\pr}_{2j,\si}\psi^{\pr}_{2j\pm1,\si})
\right]\\
+2\sinh\kappa\left[ \bar{\psi}_{2j\pm1,\uparrow}
(\bar{a}_{\pm}\psi^{\pr}_{2j\pm1,\uparrow}-
\bar{b}_{\pm}\psi_{2j,\uparrow})
(\bar{a}_{\pm}\bar{\psi}_{2j\pm1,\downarrow}-
\bar{b}_{\pm}\bar{\psi}^{\pr}_{2j,\downarrow})
\psi^{\pr}_{2j\pm1,\downarrow}\right.\nn\\
\left.\left.+\bar{\psi}^{\pr}_{2j,\uparrow}
(\bar{a}_{\pm}\psi_{2j,\uparrow}+
\bar{b}_{\pm}\psi^{\pr}_{2j\pm1,\uparrow})
(\bar{a}_{\pm}\bar{\psi}^{\pr}_{2j,\downarrow}+
\bar{b}_{\pm}\bar{\psi}_{2j\pm1,\downarrow})\psi_{2j,\downarrow}\right]
\right\}.\nn
\end{eqnarray}

For simplicity we have written here the expression only for the case\\ 
$a_{\pm1}=a_{\pm2}=a_{\pm},\;b_{\pm1}=b_{\pm2}=b_{\pm}$
(this will not damage the $CCM$) and
$\bar{a}_{+}=e^{-\kappa/2}t_{43},\;\bar{a}_{-}=e^{-\kappa/2}t_{34},\;
\bar{b}_{+}=e^{-\kappa/2}t_{41},\;\bar{b}_{+}=e^{-\kappa/2}t_{32}$
are the average values of hopping parameters.

The fermionic fields $\psi_{\uparrow}$ and $\psi_{\downarrow}$
in the expression (\ref{Rhub}) appeared because the operators 
$R$ and $R^+$ 
in the direct products in (\ref{ZZR}) are acting on independent spaces
and we should introduce different coherent fields for them.

What is left now to say, that the (\ref{Rhub}) is the expression
for the $R$-operator of the generalization of the Hubbard model
\begin{equation}
\label{RShastry}
\check{R}_{12}=e^{-h_L(u)(2n_{1,\uparrow}-1)(2n_{1,\downarrow}-1)}
\check{R}^{XX}_{12,\uparrow}\check{R}^{XX}_{12,\downarrow}
e^{-h_R(u)(2n_{2,\uparrow}-1)(2n_{2,\downarrow}-1)}
\end{equation}
with the condition $h_L(u)=h_R(u)=\kappa(u)/4$, written in the basis of
coherent states, as it was described above.

It is necessary now to mention two things:

a)The $R$-matrix of ordinary Hubbard model contained the exponent in 
(\ref{RShastry}), which is responsible for the interaction,
only in the right(or left) hand sides of the product of
two $XX$ models $R$-matrices with $\uparrow$ and $\downarrow$
spins \cite{Sh, OW, W};

b)Averaging the $CCM$ we have obtained Hubbard type model
with staggered disposition of $R$-matrices. A similar type of integrable
model is developed in \cite{ASSS2}.

\section{The action on $ML$ for any model described by $R$-matrix}

It is not hard to realize now, that the formulated above technique
is quite general and allows to pass from Hamiltonian to the Action
(Transfer matrix) formalism for any 2D model, which has a description
via $R$-matrix. Let $R_{\alpha\gamma}^{\alpha^{\pr}\gamma^{\pr}},
\;\;\alpha,\;\gamma=1,...l$ is the $R$-matrix of some model, which
has $l$-degrees of freedom at the sites of the chain.

In a beginning we should fermionize the model (see \cite{AKMS} for
details) by considering Fock space of $r$-scalar fermions
(with $l\leq 2^r$ ) $c^+_{i,s},\;c_{i,s},\; s=1,...r$ at each site
$i$ of the chain with basis
\begin{equation}
\label{nn}
\mid
n_{i,1},...n_{i,r}\rangle=c^{+n_1}_{i,1}...c^{+n_r}_{i,r}|0\rangle,
\end{equation}
and restrict the appearance of the ($2^r-l$) basic states by
applying with appropriate projectors on them as
\begin{eqnarray}
\label{mn}
\mid\alpha_i\rangle
= \Delta_1...\Delta_{2^r-l}\mid n_{i,1},...n_{i,r}\rangle.
\end{eqnarray}

As an example one can mention the 3-state $t-J$ model of two
fermions (spin $\uparrow$ and spin $\downarrow$) with restriction
on double occupancy ($\Delta_1=(1-n_{\uparrow}n_{\downarrow})$).

Let us define now the fermionic $R$-operator
\begin{equation}
\label{rxx}
\check{R}_{ij}=\check{R}_{ij,\alpha\gamma}^{\alpha^{\pr}\gamma^{\pr}}
X_{i,\alpha^{\pr}}^{\alpha} X_{j,\gamma^{\pr}}^{\gamma}
(-1)^{p(\alpha)p(\gamma^{\pr})},
\end{equation}
where $X_{i,\alpha^{\pr}}^{\alpha}=|\alpha\rangle\langle\alpha^{\pr}|$
is the Hubbard operator and $p(\alpha)$ is the fermionic parity
of the state $|\alpha\rangle$.

One can now consider the coherent states for the all $r$-copies
of fermions, extend the definition (\ref{Rpsi})
for the $R$-operator and express it as an exponent of some action
term, written for the $B$-plaquettes
\begin{eqnarray}
\check{R}_{\{\psi_{j,r}\},\{\bar{\psi}_{j+1,r}\}}^
{\{\bar{\psi}^{\pr}_{j,r}\},\{\psi^{\pr}_{j+1,r}\}}=
\langle\{\bar{\psi}^{\pr}_{j,r}\},\{\psi^{\pr}_{j+1,r}\}\mid\check
R\mid \{\psi_{j,r}\},\{\bar{\psi}_{j+1,r}\}\rangle\nn\\
\label{Rpsi2}
=exp\left\{-\cS_{j,r}(\{\psi_{j,r}\},\{\bar{\psi}_{j+1,r}\},
\{\bar{\psi}^{\pr}_{j,r}\},\{\psi^{\pr}_{j+1,r}\})\right\}.
\end{eqnarray}
Then the full action of the model will be
\begin{equation}
\label{s}
\cS=\prod_{B-plaquettes}\cS_{j,r}(\{\psi_{j,r}\},\{\bar{\psi}_{j+1,r}\},
\{\bar{\psi}^{\pr}_{j,r}\},\{\psi^{\pr}_{j+1,r}\})+\sum_{j,s}
\bar{\psi}_{j,s}\psi_{j,s}.
\end{equation} 

\section{Integrable $\cU_q(gl(n))$ models with staggered disposition of
  $R$-matrices} 

In this section we will present the main results \cite{ASSS1, APSS,AASSS, TS}
of construction
of integrable models with staggered disposition of $R$-matrices along
chain and time directions.\footnote{This results are obtained
in collaboration with J.Ambjorn, D.Arnaudon, R.Poghossian,T.Sedrakyan 
and P.Sorba}

Let us consider now $\ZZ_2$ graded quantum $V_{j,\rho}(v)$ 
(with $j=1,.....N$ as a chain index) and 
auxiliary $V_{a,\si}(u)$ spaces, where $\rho, \si =0,1$ are
the grading indices. Consider
$R$-matrices, which act on the direct product
of  spaces $V_{a,\si}(u)$ and $ V_{j,\rho}(v)$, $(\si,\rho =0,1)$,
mapping them on the intertwined direct product of 
$V_{a,\bar{\si}}(u)$ and $ V_{j,\bar{\rho}(v)}$ with the complementary
$\bar{\si}=(1-\si)$, $\bar{\rho}=(1-\rho)$ indices
\begin{eqnarray}
\label{R1}
R_{aj,\si \rho}\left( u,v\right):\quad V_{a,\si}(u)\otimes 
V_{j,\rho}(v)\rightarrow V_{j,\bar{\rho}}(v)\otimes V_{a,\bar{\si}}(u).  
\end{eqnarray}

It is convenient to introduce 
two transmutation operations $\iota_1$
and $\iota_2$ with the property $\iota_1^2=\iota_2^2=id$ 
for the quantum and auxiliary spaces
correspondingly, and to mark the operators $R_{aj,\si\rho}$ as 
follows
\begin{eqnarray}
R_{aj,00}&\equiv& R_{aj},\qquad R_{aj,01}\equiv R_{aj}^{\iota_1},\nn\\
\label{R2}
R_{aj,10}&\equiv& R_{aj}^{\iota_2},\qquad R_{aj,11}\equiv R_{aj}^{\iota_1 
\iota_2}.
\end{eqnarray}

The introduction of the $\ZZ_2$ grading of quantum spaces 
in time direction means, 
that we have now two Monodromy operators $T_{\rho}, \rho=0,1$,
which act on the space $V_{\rho}(u)=\prod_{j=1}^N V_{j,\rho}(u)$
by mapping it on $V_{\bar{\rho}}(u)=\prod_{j=1}^N V_{j,\bar{\rho}}(u)$
\begin{eqnarray}
\label{T}
T_\rho(v,u) \qquad : V_\rho(u) \rightarrow V_{\bar{\rho}}(u), \qquad \qquad 
\rho=0,1.
\end{eqnarray}

It is clear now, that the Monodromy operator of the model, which is defined
by translational invariance in two steps in the time direction and  
determines the partition function, is the product of two Monodromy operators
\begin{eqnarray}
\label{TT}
T(v,u) = T_0(v,u) T_1(v,u).
\end{eqnarray}

The $\ZZ_2$ grading of auxiliary spaces along the chain direction means
that the $T_0(u,v)$ and $T_1(u,v)$ Monodromy matrices are defined
as a staggered product
of the $R_{aj}(v,u)$ and $\bar{R}_{aj}^{\iota_2}(v,u)$ matrices:
\begin{eqnarray}
T_1(v,u)=\prod_{j=1}^N R_{a,2j-1}(v,u)
\bar{R}_{a,2j}^{\iota_2}(v,u)\nn\\
\label{T1}
T_0(v,u)=\prod_{j=1}^N \bar{R}_{a,2j-1}^{\iota_1}(v,u)
R_{a,2j}^{\iota_1 \iota_2}(v,u),
\end{eqnarray}
where the notation $\bar{R}$ denotes a 
different parameterization of the $R(v,u)$-matrix via spectral
parameters
$v$ and $u$ and can be considered as an operation
over $R$ with property $\bar{\bar{R}}= R$.
For the integrable models where the intertwiner matrix $R(v-u)$
simply depends
on the difference of the spectral parameters $v$ and $u$ 
this operation means the shift of its argument $u$ as follows
\begin{equation}
\label{RR}
\bar{R}(u)=R(\bar u), \qquad \bar{u}=\zeta-u,
\end{equation}
where $\zeta$ is an additional model parameter. 

This definitions of the Monodromy matrices cam be obtained
from the disposition of $B$-plaquettes on the $ML$, when we
are considering a chains under the angle $\pi/4$ with respect
to those of (\ref{MRR}) and staggering corresponds to $CCM$. 

As it is well known in Bethe Ansatz Technique \cite{B1,F1, F2}, the sufficient
condition for the commutativity of transfer matrices $\tau(u)=
Tr T(u)$ with different spectral parameters is the YBE. For our
case we have a two sets of equations \cite{APSS}
\begin{eqnarray}
  \label{eq:YBE1}
  R_{12}(u,v) \bar{R}_{13}^{\iota_1}(u) R_{23}(v)=
  R_{23}^{\iota_1}(v) \bar{R}_{13}(u) \tilde{R}_{12}(u,v)
\end{eqnarray}
\begin{eqnarray}
  \label{eq:YBE2}
  \tilde{R}_{12}(u,v) R_{13}^{\iota_1 \iota_2}(u) 
  \bar{R}_{23}^{\iota_2}(v)=
  \bar{R}_{23}^{\iota_1 \iota_2}(v) R_{13}^{\iota_2}(u) R_{12}(u,v) \;,
\end{eqnarray}
with $\bar{R}(u)\equiv R(\bar{u})$ and $R^{\iota_2}(u)=R^{\iota_1}(-u)$.

From $R(u)$ above, we  follow a procedure which is the inverse of
the Baxterisation (debaxterisation) \cite{Jones}.  
Let 
\begin{equation}
  \label{eq:deBaxterise}
  R_{12}(u) = \frac{1}{2i} \left(z R_{12} - z^{-1} R_{21}^{-1} \right)
\end{equation}
with $z=e^{iu}$ and the constant $R_{12}$ and $R_{21}^{-1}$ matrices are
spectral parameter independent. Then the Yang--Baxter equations 
(\ref{eq:YBE1})--(\ref{eq:YBE2})
for the spectral parameter dependent $R$-matrix $R(u)$ and
$R^{\iota_1}(u)$ are
equivalent to the following equations for the constant $R$-matrices

\begin{eqnarray}
  \label{eq:YBEconst1}
  R_{12} R_{13}^{\iota_1} R_{23} &=&
  R_{23}^{\iota_1} R_{13} R_{12}^{\iota_1} \\
  \label{eq:YBEconst2}
  R_{12}^{\iota_1} R_{13} R_{23}^{\iota_1} &=&
  R_{23} R_{13}^{\iota_1} R_{12} \\
  \label{eq:YBEconst3}
  R_{12} \left(R_{31}^{\iota_1}\right)^{-1} R_{23} 
  &-& \left(R_{21}\right)^{-1} R_{13}^{\iota_1}
  \left(R_{32}\right)^{-1}=\\ 
&=&R_{23}^{\iota_1} \left(R_{31}\right)^{-1} R_{12}^{\iota_1} 
  - \left(R_{32}^{\iota_1}\right)^{-1} R_{13}
  \left(R_{21}^{\iota_1}\right)^{-1}\nn\\
  \label{eq:YBEconst4}
  R_{12}^{\iota_1} \left(R_{31}\right)^{-1} R_{23}^{\iota_1} 
  &-& \left(R_{21}^{\iota_1}\right)^{-1} R_{13}
  \left(R_{32}^{\iota_1}\right)^{-1}=\\
&=& R_{23} \left(R_{31}^{\iota_1}\right)^{-1} R_{12} 
  - \left(R_{32}\right)^{-1} R_{13}^{\iota_1} \left(R_{21}\right)^{-1}\nn 
\end{eqnarray}
assuming $\tilde{R} = R^{\iota_1}$.

If this modified $YBE$'s have a solution, then one can formulate
a new integrable model on the basis of existing ones. 
It appeared that in the $gl(N)$ case
the two constant $R$-matrices $R$ and $R^{\iota_1}$ given by 
\begin{equation}
  \label{eq:Rsln}
  R = \sum_{i=1}^{N} q e_{ii} \otimes e_{ii}
  + \sum_{i,j=1 \atop i\neq j}^{N}  e_{ii} \otimes e_{jj}
  + (q-q^{-1}) \sum_{i,j=1 \atop i>j}^{N} e_{ij} \otimes e_{ji}
\end{equation}
\begin{equation}
  \label{eq:Rslniota1}
  R^{\iota_1} = \sum_{i=1}^{N} q e_{ii} \otimes e_{ii}
  + \sum_{i,j=1 \atop i\neq j}^{N} b_{ij} e_{ii} \otimes e_{jj}
  + (q-q^{-1}) \sum_{i,j=1 \atop i>j}^{N} e_{ij} \otimes e_{ji}
\end{equation}
satisfy the four equations (\ref{eq:YBEconst1})--(\ref{eq:YBEconst4})
provided that  
$\;\;b_{ij} = b_{ik} b_{kj}\;\; \mbox{and} \;\;b_{ij}^2=1$

By construction, all this models are of the ladder type.

\begin{acknowledgments}
The author wish to thank his
colleagues J.Ambjorn D.Arnaudon, H.Babujian, A.Belavin, R.Flume, T. Hakobyan, 
D.Karakhanyan, R.Poghossian, V.Rittenberg, T. Sedrakyan, 
P.Sorba for numerous and productive discussions.
\end{acknowledgments}

\end{document}